\documentclass[floats,floatfix,showpacs,preprintnumbers,amssymb,prd,twocolumn,superscriptaddress,nofootinbib,nolongbibliography,reprint]{revtex4-2}
\usepackage{amssymb,amsmath,verbatim,mathtools,needspace,enumitem,etoolbox,graphicx,physics,microtype,afterpage,xspace,tabularx,lmodern,multirow}
\usepackage{gensymb}
\usepackage[normalem]{ulem}
\usepackage[dvipsnames, usenames]{xcolor}
\usepackage{xr-hyper}
\definecolor{linkcolor}{rgb}{0.0,0.3,0.5}
\usepackage[unicode, colorlinks=true, linkcolor=linkcolor, citecolor=linkcolor, filecolor=linkcolor, urlcolor=linkcolor, linktocpage, breaklinks]{hyperref}
\usepackage[all]{hypcap}
\usepackage[T1]{fontenc}
\usepackage[utf8]{inputenc}
\usepackage[export]{adjustbox}
\usepackage[usenames,dvipsnames]{xcolor}
\hypersetup{colorlinks=true,citecolor=romared,linkcolor=romared,urlcolor=romared}

\definecolor{romared}{RGB}{142,0,28}

\newcommand{\be}{\begin{equation}}
	\newcommand{\ee}{\end{equation}}

\def\be{\begin{equation}}
	\def\ee{\end{equation}}
\newcommand{\beq}{\begin{eqnarray}}
	\newcommand{\eeq}{\end{eqnarray}}

\usepackage{makecell}
\usepackage{soul}

\newcolumntype{Y}{>{\centering\arraybackslash}X}

\definecolor{romared}{RGB}{142,0,28}

\makeatletter
\newcommand*{\addFileDependency}[1]{
	\typeout{(#1)}
	\@addtofilelist{#1}
	\IfFileExists{#1}{}{\typeout{No file #1.}}
}
\makeatother

\begin{document}
	\title{Hushing black holes: tails in dynamical spacetimes}
	
	\author{Vitor Cardoso} 
	\affiliation{Niels Bohr International Academy, Niels Bohr Institute, Blegdamsvej 17, 2100 Copenhagen, Denmark}
	\affiliation{CENTRA, Departamento de F\'{\i}sica, Instituto Superior T\'ecnico -- IST, Universidade de Lisboa -- UL, Avenida Rovisco Pais 1, 1049-001 Lisboa, Portugal}
	\affiliation{Yukawa Institute for Theoretical Physics, Kyoto University, Kyoto
		606-8502, Japan}
	\author{Gregorio Carullo} 
	\affiliation{Niels Bohr International Academy, Niels Bohr Institute, Blegdamsvej 17, 2100 Copenhagen, Denmark}
	\author{Marina De Amicis} 
	\affiliation{Niels Bohr International Academy, Niels Bohr Institute, Blegdamsvej 17, 2100 Copenhagen, Denmark}
	\author{Francisco Duque}
	\affiliation{Max Planck Institute for Gravitational Physics (Albert Einstein Institute) \\
		Am Mühlenberg 1, D-14476 Potsdam, Germany}
	\author{Takuya Katagiri}
	\affiliation{Niels Bohr International Academy, Niels Bohr Institute, Blegdamsvej 17, 2100 Copenhagen, Denmark}
	\author{David Pere\~niguez}
	\affiliation{Niels Bohr International Academy, Niels Bohr Institute, Blegdamsvej 17, 2100 Copenhagen, Denmark}
	\author{Jaime Redondo-Yuste} 
	\affiliation{Niels Bohr International Academy, Niels Bohr Institute, Blegdamsvej 17, 2100 Copenhagen, Denmark}
	\author{Thomas F.M.~Spieksma} 
	\affiliation{Niels Bohr International Academy, Niels Bohr Institute, Blegdamsvej 17, 2100 Copenhagen, Denmark}
	\author{Zhen Zhong}
	\affiliation{CENTRA, Departamento de F\'{\i}sica, Instituto Superior T\'ecnico -- IST, Universidade de Lisboa -- UL, Avenida Rovisco Pais 1, 1049-001 Lisboa, Portugal}
	
	\begin{abstract}
		Stationary, asymptotically flat, black hole solutions of the vacuum field equations of General Relativity belong to the Kerr family. But how does one approach this state, dynamically? Linearized fluctuations decay at late times, at fixed spatial position, as a Price power law for generic initial conditions. However, little attention was paid to forced and nonlinear spacetimes, where matter and nonlinearities play a role. We uncover a new, source-driven tail governing waves generated by pointlike matter and nonlinearities, which can dominate over Price's decay.
	\end{abstract}
	
	\maketitle
	
	\noindent {\bf \em Introduction.} 
	General Relativity is our ultimate description of the gravitational interaction. It contains, as a mathematical solution, perhaps the most interesting and bizarre object known to humankind: black holes (BHs)~\cite{Hawking:1987en,Misner:1973prb,Chandrasekhar:1985kt}. Two powerful results highlight the role of BHs in our understanding of nature and on their potential to uncover new laws of physics: i. the most general vacuum BH solution belongs to the Kerr family~\cite{Carter:1971zc,Robinson:1975bv,Chrusciel:2012jk}; ii. BHs harbour spacetime singularities, where the known laws of physics break down~\cite{Penrose:1964wq,Penrose:1969pc}. The assertion that all singularities are hidden from us is so remarkable that any observational evidence for or against it is highly sought for. 
	
	The advent of gravitational-wave astronomy harnessed access to information from the coalescence of two BHs, and opened new exciting ways to test gravity in its strong-field and dynamical regime. One of the relevant questions to ask concerns the approach to the final state: if there is a stationary state and it is well described by the Kerr family, how is it approached? How does the spacetime shed its multipolar structure when approaching a stationary BH state? Insight into this problem was provided by linearizing the field equations. At late times, BHs relax (``ring down'') in a series of exponentially damped sinusoids, so-called quasinormal modes~\cite{Berti:2009kk}, followed by an inverse power-law decay with time, Price ``tails''~\cite{Price:1971fb,Leaver:1986gd,Gundlach:1993tp,Hintz:2020roc}. 
	For generic initial data of multipolar structure described by a spherical harmonic of angular number $\ell$, a massless field in a non-spinning BH background decays at fixed spatial position as $\Phi\sim t^{-p},\,p=2\ell+3$ at very late times~\cite{Price:1971fb}. Such behavior arises from a branch-cut in the Green's function, or equivalently from large-radius scattering, and is observed in linearized analysis of scattering experiments, stellar collapse or point particle evolutions in BH backgrounds~\cite{Gundlach:1993tp,Cardoso:2003jf,Harms:2014dqa}. We term such decay ``Initial Data-led Tails'' (IDT). In numerical experiments, power-law tails are suppressed relative to the ringdown stage, hence they are very challenging to observe. Nevertheless, they dictate how fluctuations behave at late times and play a crucial role in foundational issues such as strong cosmic censorship~\cite{Cardoso:2017soq,Hod:2018dpx,Cardoso:2018nvb,Reall_CCC}.
	
	It has recently been observed in linear and nonlinear simulations of eccentric coalescences (see Fig. 19 in Ref.~\cite{Albanesi:2023bgi} and Fig. 1 in Ref.~\cite{Carullo:2023tff}) that the ringdown stage can be much shorter than previously thought, and very soon dominated by large amplitude transients which decay slower than IDTs (reporting fall-offs at null infinity slower than the asymptotic Price value at intermediate times). This behavior is unexpected, and exciting, as it means that large-amplitude post-ringdown signals might be detected with upcoming experiments, leading to new tests of Einstein equations with gravitational waves. If such a decay really describes the long-term behavior of relaxing BH spacetimes, it also means that our understanding of final state problem (how does one get to the final Kerr state) is incomplete; should such a decay also be present in cosmological backgrounds, then there are also important consequences for strong cosmic censorship.
	
	We see three possibilities to circumvent Price's results: a) the effect is linear but triggered by the motion of sources in the BH vicinity, something not taken into account in all previous asymptotic analyses, which focused on the vacuum (and sourceless) linearized equations. As we will show, source-driven tails---SoDTs---do indeed exist, and depend on the source spatial distribution; b) the effect is nonlinear in nature. Such a possibility was raised in the past in the context of toy models~\cite{Okuzumi:2008ej}. Here we show, in the context of second-order perturbation theory, that second order tails exist and may dominate over linear ones, yet are weaker than predicted~\cite{Okuzumi:2008ej}; c) the effect is only a transient, possibly triggered by initial conditions. 
	
	\noindent {\bf \em Initial and source-driven tails.} 
	Building on previous work~\cite{Barack:1998bw,Karkowski:2003et,Price:2004mm}, we develop a two-parameter perturbative expansion to estimate the asymptotic decay of the fields in the presence of sources. Details are given in Supplemental Material (SM). Consider the wave-like equation describing massless field evolutions in a spherically symmetric BH spacetime, written in double null coordinates~$(u,v)$,
	\begin{equation}
		-4\partial^2_{uv}\Phi = V\Phi +\alpha\mathcal{S}\,,\label{EquationforPhi}
	\end{equation}
	where $V=V(x)$ is an effective potential that depends only on the tortoise coordinate $x\in ]-\infty,+\infty[$; $\alpha$ is a book-keeping parameter that controls the source~$\mathcal{S}$. Since the background has a Killing vector~$T = (1/2)(\partial_u + \partial_v)$, the potential is a function of $x = (v-u)/2$~\cite{Gundlach:1993tp}. We decompose the potential into a truncated centrifugal potential barrier,~$V_0$, and a perturbation~$\delta V$,
	\begin{equation}
		V = V_0+\varepsilon \delta V\,,
	\end{equation}
	with
	\begin{equation}
		V_0 = \begin{cases}
			\dfrac{\ell(\ell+1)}{x^2}\,, \quad &x \geq x_0 \,  \\
			0 \,, \quad &x < x_0
		\end{cases} \,.
	\end{equation}
	Here, $x_0$ characterizes wave scattering and is of the order of the light ring radius, whereas $\varepsilon$ is a small book-keeping parameter. We consider\footnote{In a spacetime of mass $M$, one can expand $r=x-2M\log{x}+4M^2/x(1+\log{x})+...$ . For massless fields, $\rho=3$. One can extend the analysis to $\delta V = x^{-\rho}\log x$~\cite{Ching:1994bd,Ching:1995tj}. }
	\begin{equation}
		\label{eq:deltaV}
		\delta V = x^{-\rho}\,,
	\end{equation}
	with $\rho=3,4,5,\cdots$. The correction~$\delta V$ and the source~${\cal S}$ are responsible for the appearance of late-time tails, which we now study. It is known that a branch cut in the Green's function in the complex frequency plane is a generic feature if $\delta V$ tends to zero asymptotically slower than an exponential~\cite{Ching:1994bd,Ching:1995tj}. 
	Consider $\Phi$ in a two-parameter perturbative expansion,
	\begin{equation}
		\Phi = \Phi^{(0,0)} + \varepsilon\Phi^{(1,0)} + \alpha\Phi^{(0,1)} + \alpha\varepsilon\Phi^{(1,1)} + \dots\,.
		\label{eq:twoparameterExpforPhi}
	\end{equation}
	For~$\alpha=0$ this expansion gives a good description of the late-time behavior~\cite{Price:2004mm}. 
	Clearly, $\Phi^{(0,0)}$ contains the direct propagation of the initial conditions through the centrifugal barrier (flat space, if $\ell = 0$).
	
	In summary, we find the following. In the absence of a source (see also footnote~\ref{footnote:remark})
	\beq
	\Phi_{\cal I^+}^{(1,0)}(u) \sim u^{1-\rho-\ell}\,,\qquad
	\Phi^{(1,0)} \sim t^{-\rho-2\ell}\,,~~t\gg x\,. \label{eq:IDTs}
	\eeq
	For $\rho=3$, we recover Price's law~\cite{Price:1971fb}.\footnote{\label{footnote:remark} We are mostly interested in waves at $t \gg r={\rm constant}$ and large. One can also use null coordinates and focus on ${\cal I}^+$ ($v\to \infty$ at constant $u$). For $\Phi\sim t^{-m}$ then $\Phi\sim u^{-m+\ell+1}$ at ${\cal I}^+$~\cite{Price:1971fb,Barack:1998bw}.
	}
	
	Sources of the form (or their extended version, see SM)
	\begin{equation}
		\label{localisedsources}
		\mathcal{S} = \frac{\delta(x-x_s-U_s t)}{x^\beta}\,,~~\beta\ge 0\,,
	\end{equation}
	where $U_s$ ($|U_s|\le 1$) is a source velocity, are physically interesting, as they describe compact objects moving in the vicinity of BHs, plunging ($U_s<0$) or moving outwards ($U_s>0$). Outward motion is a good prototype for gamma-ray bursts~\cite{Birnholtz:2013bea}, or to mimic eccentric motion where long-lived transients have been observed~\cite{Albanesi:2023bgi,Carullo:2023tff}.
	We find, 
	
	\noindent {\bf i)}~for outward-directed sources and $\beta \le \ell+2$ with $0<U_s< 1$, SoDTs of the form
	\begin{equation}
		\Phi_{{\cal I^+},0<U_s< 1}^{(0,1)}\sim u^{1-\beta}\,,\quad 
		\Phi_{0<U_s<1}^{(0,1)}\sim t^{-\beta-\ell},~~t\gg x\,.\label{eq:LSoDTs}
	\end{equation}

	\noindent {\bf ii)}~for outward-directed sources and $\beta\ge \ell+3$ with $0<U_s\le 1$, SoDTs~\eqref{eq:LSoDTs} are generated but the late-time behavior is dominated by the IDTs, Eq.~\eqref{eq:IDTs}.
	
	\noindent {\bf iii)}~for outward-directed sources of $\beta=0,1$ or $\beta=\ell+2$ with $U_s=1$, SoDTs of the form
	\begin{equation}
		\Phi_{{\cal I^+},U_s=1}^{(0,1)}\sim u^{1-\beta}\,,\quad 
		\Phi_{U_s=1}^{(0,1)}\sim t^{-\beta-\ell},~~t\gg x\,,\label{eq:LSoDTsNull}
	\end{equation}
	but when $2\le \beta\le \ell+1$ for $\ell\ge1$,
	\begin{equation}
		\Phi_{{\cal I^+},U_s=1}^{(0,1)}\sim u^{-\beta}\,,\quad 
		\Phi_{U_s=1}^{(0,1)}\sim t^{-\beta-\ell-1},~~t\gg x\,.\label{eq:LSoDTsNull2}
	\end{equation}
	
	\noindent {\bf iv)}~ in the presence of the inward-directed sources with $-1\le U_s<0$, no SoDTs appear at late times. The late-time behavior is dominated by IDTs~\eqref{eq:IDTs}.
	
	Problem \eqref{EquationforPhi} can describe a linearized setting with pointlike masses, or a nonlinear problem expanded to second order. Our analytical results therefore predict a nontrivial late-time decay, which can be source-dominated. The following explores the same problem, but from a numerical perspective, confirming our predictions, both for pointlike sources and for sources appropriate for the second-order expansion of the Einstein equations, thus establising our results in the nonlinear regime. Details on the numerical procedure are given in the SM and Refs.~\cite{Krivan_1997, Pazos_valos_2005, Zenginoglu:2011zz, Zenginoglu:2012us,Calabrese:2005fp,Lopez-Aleman:2003sib}.
	
	\noindent {\bf \em Numerical results I: pointlike objects around a BH.} 
	%
	\begin{figure}[h!]
		\includegraphics[width=\linewidth]{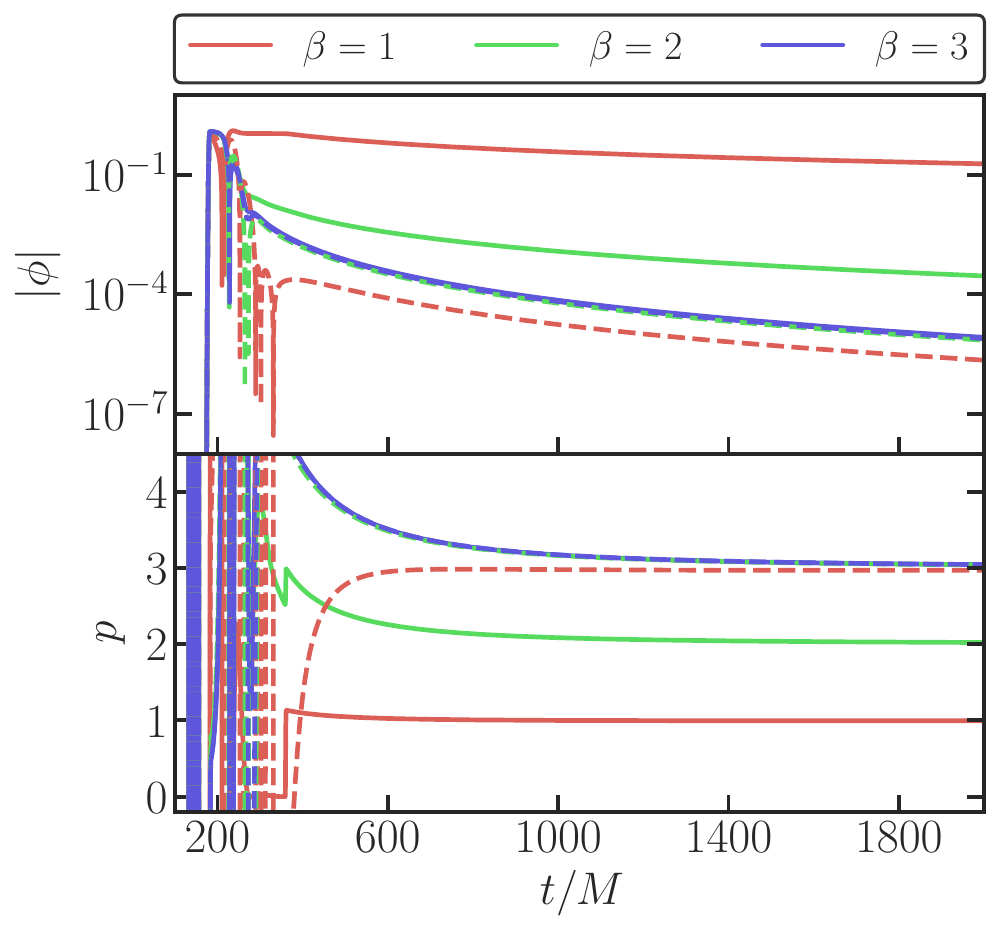}
		\caption{Scattering of a scalar field ($\ell=0$) in a Schwarzschild background, Eq.~\eqref{eq:kg_schw_wave2} with $U_s = \pm 0.5$ and non-static initial data $\partial_t\Phi\neq0$ (see SM). Source is initialized at $x_s = 20M$, $\phi$ is extracted at $x_{\rm ext} = 200M$. Solid (dashed) lines denote outward (inward) motion. The power-law is independent on initial data for $U_s < 1$, and agrees with prediction~\eqref{eq:LSoDTs}. Throughout this work, the exponent $p$ is calculated via $p=-t\partial_t \ln|\phi|$. Steep jumps in outward motion occur after source passage (see SM).
			\label{fig:Schwarzschild_gaussion_v0.5}
		}
	\end{figure}
	Consider now a $3+1$ problem on a BH background. We focus on a scalar field, sourced by a pointlike charge of the form:
	\begin{equation}
		\nabla^a \nabla_a \Phi = \sum_{\ell, m}\frac{\delta(x-x_s-U_st)}{r^{\beta+1}} Y_{\ell m}\,, \label{eq:kg_schw_gaussian}
	\end{equation}
	where $Y_{\ell m}$ are standard spherical harmonics. We take the ansatz $\Phi = \phi(t, r) Y_{\ell m}/r$ and find,
	\begin{equation}
		\partial^{2}_t\phi - \partial^{2}_x\phi + f\frac{r f' + \ell (\ell + 1)}{r^2}\phi + \frac{f \delta(x-x_s - U_st)}{r^{\beta}} = 0\,, \label{eq:kg_schw_wave2}
	\end{equation}
	with $f=1-2M/r$ and the tortoise coordinate $x$ defined in terms of the coordinate $r$ via $\mathrm{d}r/\mathrm{d}x=f$. The peculiar choice of the power $(\beta+1)$ in \eqref{eq:kg_schw_gaussian} is made such that $\beta$ in \eqref{eq:kg_schw_wave2} is the curved space analog of the 1+1 source \eqref{localisedsources}.
	
	As apparent in Fig.~\ref{fig:Schwarzschild_gaussion_v0.5}, inward-moving sources excite BH ringdown, a stage which subsequently gives rise to Price tails~\eqref{eq:IDTs}. On the other hand, outward-moving sources give rise to SoDTs \eqref{eq:LSoDTs}, which are longer lived (for $\beta<\ell+3$) than Price tails. The results concern $U_s=\pm 0.5$, but the outcome is qualitatively the same for other velocities and for Gaussian widths which are orders of magnitude larger. At late times and for $|U_s|<1$, we obtain at finite radii~\cite{comment_exception} 
	\beq
	\Phi &\sim & t^{-\beta-\ell}\,,\quad 0<U_s<1 \quad\& \quad \beta \leq \ell + 2\,,\nonumber\\
	\Phi &\sim & t^{-3-2\ell} \,,\quad {\rm otherwise} \label{eq:finitev}
	\eeq
	in perfect agreement with the analytical predictions~\eqref{eq:LSoDTs}. All the results we discuss were also measured at null infinity to obey the relation in footnote~\ref{footnote:remark}. For $\beta \geq \ell + 3$, SoDTs are also excited (according to our analytical results) but are subdominant with respect to IDTs. In agreement with analytical predictions, we find that inwards-directed sources only excite IDTs. It is worth mentioning that {\it before} the particle passes the observer, for some regimes---specially low $U_s$---we find still a power-law behavior, but now of the form,
	$\Phi \sim t^{1-\beta+\ell}\,,\,{\rm for\,\,\,} t<(x-x_s)/U_s$.
	
	The above results are independent of velocity, for $|U_s|<1$, and of initial conditions. For an {\it inwards} travelling source with $U_s = -1$, we find Price tails in all cases, i.e.
	$\Phi\sim t^{-3-2\ell}$. 
	On the other hand, for an {\it outwards} travelling source~($U_s = 1$)~\cite{comment_exception}
	\beq
	\Phi &\sim & t^{-\beta-\ell}\,, \quad {\rm for} \quad  \beta= 0, 1\,,\nonumber\\
	\Phi &\sim&  t^{-2-2\ell}\,, 
	\quad  {\rm for } \quad 2\le \beta \leq \ell +2\,,\nonumber\\
	\Phi &\sim & t^{-3-2\ell}\,, 
	\quad  {\rm otherwise }\,.\label{eq:Static_v1}
	\eeq
	(the power $p$ at intermediate times is sensitive to the initial location $x_s$ of the source and to its width; we start the particle close to the BH ($x_s<0$) to suppress ``junk radiation'' which could excite IDTs stronger than SoDTs).
	
	The empirical, numerical result disagrees with analytical predictions~\eqref{eq:LSoDTsNull}--\eqref{eq:LSoDTsNull2} for $2\le \beta \le \ell+2$. We attribute this to the fact that $\Phi^{(0,1)}$ relies on the perturbative expansion~\eqref{eq:twoparameterExpforPhi}, not the exact problem~\eqref{eq:kg_schw_wave2}, missing some properties of $\Phi$. 
	
	\begin{figure}
		\includegraphics[width=\linewidth]{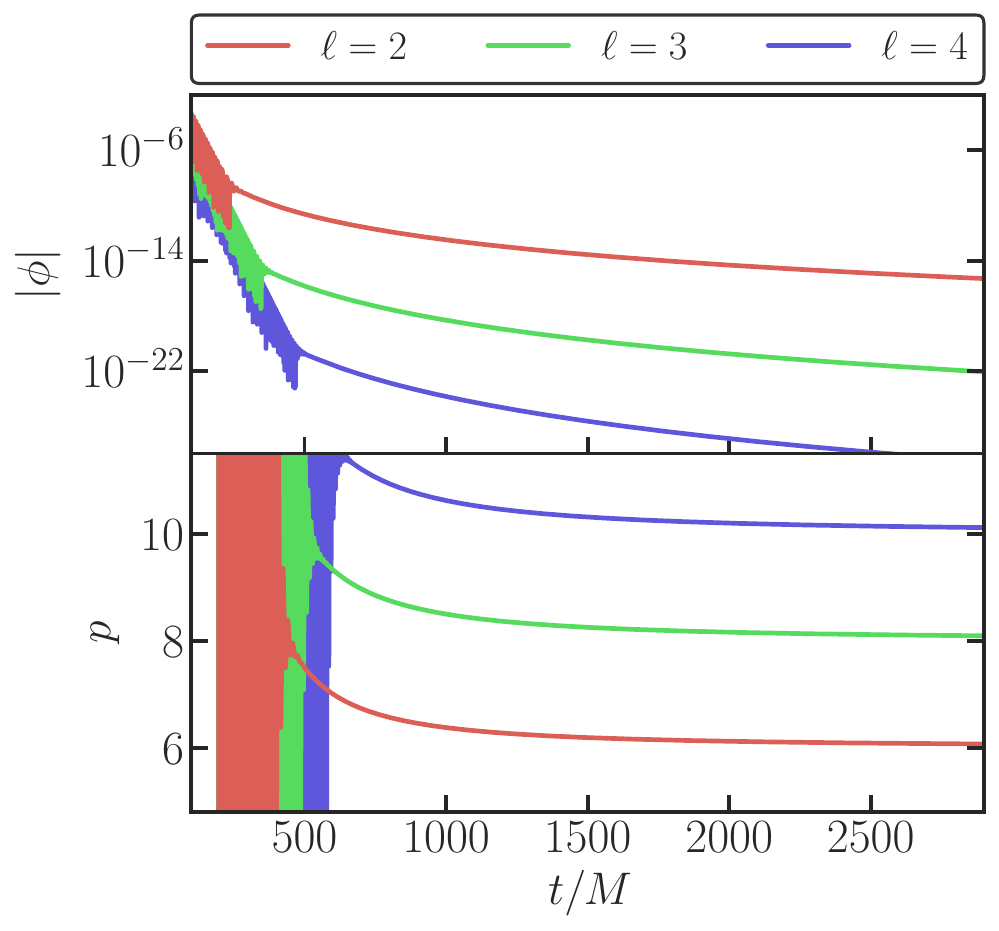}
		\caption{Tails for massless particle on outgoing radial geodesics on a Schwarzschild background. We use $\Phi (0, x) = \partial_t \Phi (0, x) = 0$. Particle is initialized at $x_s = -20M$, $\phi$ extracted at $x_{\rm ext} = 200M$. The asymptotic behavior agrees with~\eqref{eq:Static_v1} for $\beta=2$.
		}
		\label{fig:null_n}
	\end{figure}
	We can repeat the numerical experiment with a realistic setup: a pointlike mass following a radial geodesic on Schwarzschild background. The particle sources GWs which are governed by the Zerilli equation (SM and Refs.~\cite{Davis:1971gg,Cardoso:2021vjq,Cardoso:2002ay}). We use vanishing initial data. Therefore, we expect to recover---and we do---the results of the previous section for $\beta = 0$ for massive particles~\eqref{eq:finitev}, and $\beta = 2, U_s=1$~\eqref{eq:Static_v1} for massless particles. The precise signal and power is shown for outward null motion in Fig.~\ref{fig:null_n}. Inward motion yields the expected IDTs.
	
	\noindent {\bf \em Numerical results II: second order tails.} 
	%
	\begin{figure}[h!]
		\includegraphics[width=0.95\linewidth]{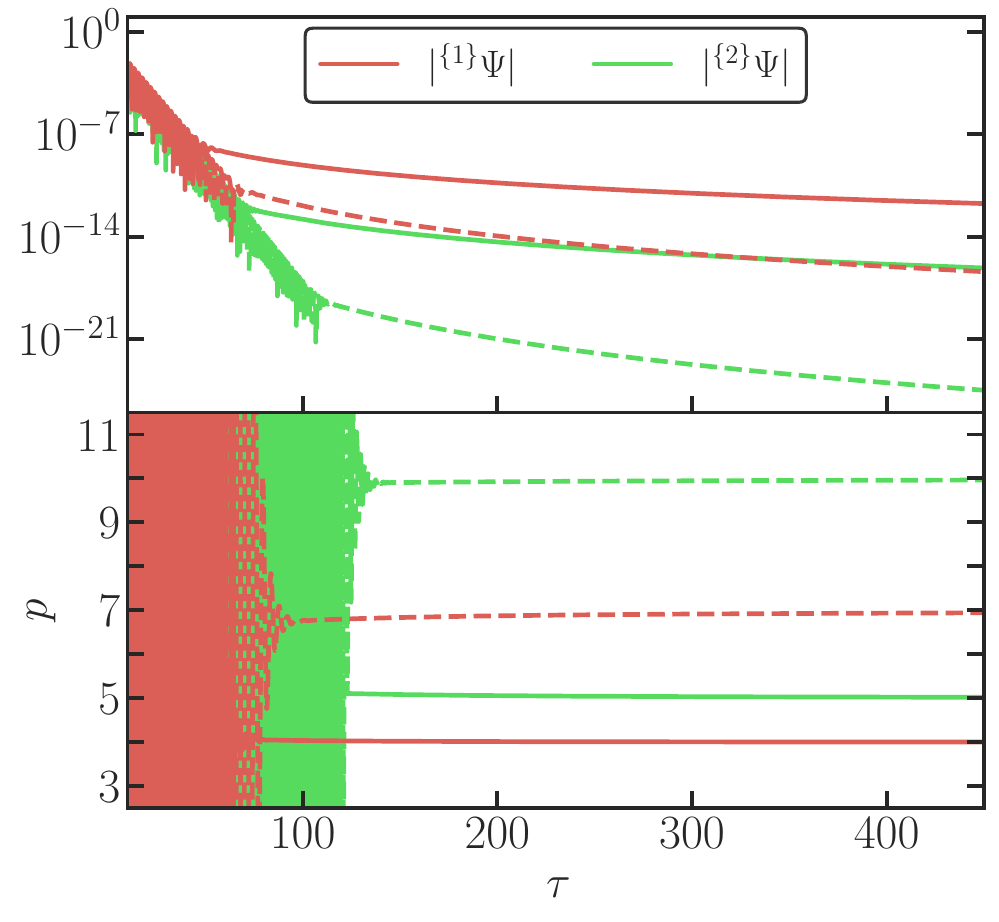}
		\caption{First-order $(\ell=m = 2)$ (red line) and second-order polar gravitational perturbations $(\ell=m=4)$ (green line) sourced by self-coupling of the former. We use sources in Ref.~\cite{Brizuela:2009qd} and evolve the equations with approximately ingoing initial data for first order perturbations, and zero initial data for second order quantities. Solid and dashed lines are extracted at future null infinity $\mathcal{I}^{+}$ and at $\rho_{\rm ext} \approx 0.1$, respectively. First order decays as Price tail, second order as~\eqref{eq:Static_v1}.
		}
		\label{fig:second_order_pert_Brizuela_ingoing0}
	\end{figure}
	Consider now the full nonlinear Einstein equations. A perturbative expansion of the field equations around a background metric can be written as $\tilde{g}_{\mu\nu}=g_{\mu\nu}+\sum_{n=1}^{\infty}\frac{\varepsilon^n}{n!}\,^{\{n\}}\hspace{-.5mm}h_{\mu\nu}$
	~\cite{Regge:1957td,Zerilli:1970se,PhysRevD.19.2268,PhysRevD.22.1300,Martel:2005ir,Gleiser:1995gx,Gleiser:1998rw,Brizuela:2006ne}. For a vacuum Schwarzschild background spacetime, $^{\{n\}}\hspace{-.5mm}h_{\mu\nu}$ can be reconstructed from the gauge-invariant generalisation of the Zerilli and Regge-Wheeler master variables~\cite{Gleiser:1995gx,Gleiser:1998rw,Brizuela:2006ne,Ioka:2007ak,Nakano:2007cj}. We focus on the Zerilli sector, governed by,
	\begin{align}
		-\partial_t^2\,^{\{2\}}
		\hspace{-.5mm}\Psi+\partial_{x}^2\,^{\{2\}}\hspace{-.5mm}\Psi-V_Z\,^{\{2\}}\hspace{-.5mm}\Psi&=\,^{\{2\}}\hspace{-.5mm}\mathcal{S}_\Psi\,,\label{eqpsi}
	\end{align}
	with the potentials given in the SM.
	
	The source term of order $n$, $^{\{n\}}\hspace{-.5mm}\mathcal{S}_\Psi$, depends on the metric perturbations up to order $n-1$, with the first order sources being identically zero~\cite{Brizuela:2006ne,Ioka:2007ak,Nakano:2007cj}.
	We focus on the expansion up to and including second order, $n=2$. Source terms decay as ${\cal O}(1)/r^2$ at infinity~\cite{Brizuela:2006ne,Nakano:2007cj}. 
	
	Evolutions of second-order perturbations were carried out in the past~\cite{Pazos:2010xf}, with a different purpose. We use a pseudospectral code, with hyperboloidal coordinates $\{\tau, \rho, \theta, \varphi\}$ in the minimal gauge \cite{Ansorg:2016ztf, PanossoMacedo:2018hab, PanossoMacedo:2018gvw}. Details on the numerical procedure and initial conditions for first order quantities are given in the SM.
	
	Results for the evolution of Eq.~\eqref{eqpsi} are shown in Fig.~\ref{fig:second_order_pert_Brizuela_ingoing0} for a $\ell=m=4$ second order perturbation driven by the self-coupling of $\ell=m=2$ quantities. With Prony methods~\cite{Berti:2007dg}, we find two families of quasinormal modes~\cite{Nakano:2007cj,Cheung:2022rbm,Mitman:2022qdl,Redondo-Yuste:2023seq}.
	The first family belongs to potential-driven modes and are also part of the spectrum of first-order quantities. Our results are compatible with the linearized prediction (to 14 decimal digits or more). The second family are source-driven, whose frequencies are twice that of the first-order perturbations~\cite{Ioka:2007ak, Nakano:2007cj,London:2014cma,Lagos:2022otp,Cheung:2022rbm,Mitman:2022qdl,Redondo-Yuste:2023seq,Khera:2023oyf}. 
	
	We find a power $p=10$ for the decay at fixed spatial distance for this particular mode, which,
	given the fall-off of the source term $\,^{\{2\}}\hspace{-.5mm}\mathcal{S}_\Psi$ at infinity ($\propto r^{-2}$), is consistent with the SoDT \eqref{eq:Static_v1}~\cite{comment_exception}. These results are to the best of our knowledge the first accurate simulations of second order gravitational modes including late-time tails. The source term in Ref.~\cite{Nakano:2007cj} yields a power $p$ which is consistently one unity less than that for the source in Ref.~\cite{Brizuela:2006ne}, since the master wavefunction of the former is, up to irrelevant factors, the time derivative of the latter. We show in the SM that for this reason there must be special classes of initial data for which the late-time decay is consistently faster than that predicted by Price's law.
	
	Our findings are consistent with those from nonlinear, spherically symmetric collapse simulations~\cite{Gundlach:1993tn,Dafermos:2003yw,Bizon:2008iz} (there are a few studies of dynamical spacetimes with no symmetries, but it is unclear whether they apply to General Relativity~\cite{Szpak:2008jv,Luk:2010gr,lindblad2022weak}).
	Our results are in tension with claims in Ref.~\cite{Okuzumi:2008ej} (also Ref.~\cite{Lagos:2022otp}), which reports a decay $t^{-2}$. To further investigate this issue without being limited by the numerical accuracy of first-order quantities, we solved the following problem with an independent numerical routine,
	\begin{equation}
		\nabla^a \nabla_a \Phi =\sum_{\ell m} \frac{e^{-i\omega (t \pm x)}H(t\pm x)}{r^{\beta+1}} Y_{\ell m}\,, \label{eq:kg_schw_ringdownsource}
	\end{equation}
	with $H$ the Heaviside function and $\omega=0.5-0.1i$ (arbitrarily chosen). This source mimics a first-order ringdown waveform. Numerical results are again consistent with our previous findings: there are two distinct tails, depending on the sign in the exponent. For $\beta<3$, we find results \eqref{eq:Static_v1} for outward-moving sources and Price's law $\Phi \sim  t^{-3-2\ell}$ for inward-moving sources (again, the analytical approach fails to capture all the features, we suspect because some properties of $\Phi$ are missing in $\Phi^{(0,1)}$).
	
	Predictions~\cite{Okuzumi:2008ej,Lagos:2022otp} of source-driven and nonlinear tails assumed that the effective potential decays asymptotically faster than the centrifugal barrier and using a (1+1)-dimensional Green's function. Such studies predicts a $1/t^2$ decay, slower than any IDT~\cite{Okuzumi:2008ej,Lagos:2022otp}. Our results are consistent with this prediction only for $\ell=0$ modes, where the centrifugal barrier is absent and indeed the Green's function is 1+1 dimensional. In such toy models, the source term decays slower than the effective potential at large distances~\cite{Okuzumi:2008ej,Lagos:2022otp}. We conjecture that this determines the asymptotic time behavior and that $p=\beta+\ell$ in such cases. 
	
	\noindent {\bf \em Discussion.} 
	We have shown that distributional sources, such as those appropriate to describe pointlike objects, or even extended coherent sources, give rise to late-time power-law tails which can dominate over initial data-driven tails. These results are important as a point of principle, but could play a role in observations in the linearized context, for example for extreme-mass-ratio systems. 
	For second-order corrections, our results establish the existence of tails which can dominate over linear ones, but which decay much faster than over-simplified toy models suggest~\cite{Okuzumi:2008ej,Lagos:2022otp}.
	
	Nonlinear corrections to extreme-mass-ratio systems will also occur, but are suppressed by the mass ratio. However, there are consequences at {\it linear} order, due to the distributional character of the source. The high-amplitude post-ringdown signal observed in extreme-mass-ratio systems and for which there is evidence in numerical evolutions of binary mergers~\cite{Albanesi:2023bgi,Carullo:2023tff} then begs the question of their origin. Since they are apparent in eccentric binaries only, we have also looked for imprints in toy models attempting to mimic eccentric binaries, such as $\nabla^a \nabla_a \Phi = \delta(x-x_s-U_st-A\cos{\Omega t})/r^{\beta+1}\,,$ but for all plunging motions the asymptotic behavior of radiation at late times is governed by the IDT $t^{-(2\ell + 3)}$. We were thus unable to reproduce the behavior reported in the literature, which requires a more specialised analysis.

	Our results, summarized in Figs.~\ref{fig:Schwarzschild_gaussion_v0.5}--\ref{fig:second_order_pert_Brizuela_ingoing0} also suggest that the on-set of the tail occurs long after the start time of ringdown for plunging particles, and that in this case their magnitudes are suppressed by orders of magnitude relative to the main burst. However, for outgoing sources, tail magnitudes are not negligible, raising questions about detectability. We also note that our results are consistent with the literature. Timelike scattering of pointlike particles was predicted to give rise to a $1/u$ leading tail~\cite{Saha:2019tub}. This result is consistent with our own numerical result for scattered particles. For ingoing sources, our results predict that the tail is suppressed, in line also with Ref.~\cite{Saha:2019tub}, which shows that if the final state is a single massive object and radiation -- as in our ingoing or outgoing massless sources -- the $1/u$ piece cancels out. This also provides consistency with Price's results and explains why tails for second order perturbations decay faster than $1/u$, because these are massless sources, for which one would expect the cancellations to occur.

	In classical mechanics and General Relativity, the future is dictated by initial conditions. The reader might then wonder why we find an asymptotic behavior different from that derived from the evolution of generic initial conditions, for linearized equations. The reason is that compact sources, such as the ones we consider, arise as the evolution of initial data only if one includes extra, nonlinear interactions, in the theory. These have not been dealt with in Price's analysis. 
	
	Finally, the existence of SoDTs is not fundamentally tied to the existence of horizons nor to the asymptotic structure of the spacetime. Hence, one might suspect they exist in cosmological backgrounds of interest, like asymptotic de Sitter spacetimes. Our results show they exist in toy models with P\"{o}schl-Teller effective potentials, which have no IDTs and which are a good proxy for Schwarzschild de Sitter spacetimes~\cite{Cardoso:2003sw}. It would be an intriguing resolution to 
	cosmic censorship violations if the very existence of matter would restore it~\cite{Cardoso:2017soq,Reall_CCC}.
	
	\noindent {\bf \em Acknowledgments.} 
	We are indebted to Jo\~ao Costa, Peter Hintz and Jos\'e Nat\'ario for very helpful comments and insights and for guidance through the mathematical physics literature. We thank Hiroyuki Nakano for providing his Maple file and for fruitful discussions on second-order perturbation theory and to David Brizuela and Manuel Tiglio for correspondence on second order problems and for graciously sharing their results and notebooks. We are also thankful to Lorenzo Annulli for discussions on the close limit approximation. 
	We acknowledge support by VILLUM Foundation (grant no. VIL37766) and the DNRF Chair program (grant no. DNRF162) by the Danish National Research Foundation.
	V.C.\ is a Villum Investigator and a DNRF Chair.  
	V.C. acknowledges financial support provided under the European Union’s H2020 ERC Advanced Grant “Black holes: gravitational engines of discovery” grant agreement no. Gravitas–101052587. 
	Views and opinions expressed are however those of the author only and do not necessarily reflect those of the European Union or the European Research Council. Neither the European Union nor the granting authority can be held responsible for them.
	This project has received funding from the European Union's Horizon 2020 research and innovation programme under the Marie Sklodowska-Curie grant agreement No 101007855 and No 101131233.
	Z.Z.\ acknowledges financial support from China Scholarship Council (No.~202106040037).
	\newpage
		
		
		\begin{center}
			{\Large \bf Supplemental material}
		\end{center}
	
	\section {A summary of known results for linear fluctuations \label{subsec:overview}} 
	Here, we provide details of our results.
	To place the context, we start by briefly summarizing the main results on the late-time decay (IDTs) of massless fields in BH backgrounds. We focus on non-spinning BHs.
	
	Linearized fluctuations of BH spacetimes seem to be well understood. For a Schwarzschild BH of mass $M$ in standard  coordinates, the evolution of these variables is governed by the wave equation~\cite{Regge:1957td,Zerilli:1970se,PhysRevD.19.2268,PhysRevD.22.1300,Martel:2005ir} ($x$ ranges from $]-\infty,+\infty[$)
	\beq
	\partial_{x}^2\Phi-\partial_t^2\Phi-V\Phi=\mathcal{S}\,.\label{eq:massless}
	\eeq
	Here $\Phi$ denotes a gauge-invariant combination of metric variables. These are expanded in tensor spherical harmonics of indices $\ell, m$ and decomposed in two different sets, so-called Regge-Wheeler or axial and Zerilli or polar. The tortoise coordinate is defined as
	\be
	x=r+2M\log(r/(2M)-1)\,,\label{tortoise_definition}
	\ee
	and the potentials $V=V_{\rm Z}$ or $V=V_{\rm RW}$ are given by
	\beq
	\begin{aligned}
		V_{\rm Z}&=f\left[\frac{\ell(\ell+1)}{r^2}-\frac{6M}{r^3}\frac{r^2\lambda(\lambda+2)+3M(r-M)}{(r\lambda+3M)^2}\right]\,,\\
		V_{\rm RW}&=f\left[\frac{\ell(\ell+1)}{r^2}-\frac{6M}{r^3}\right]\,,
	\end{aligned}
	\label{eq:potentials}
	\eeq
	where $\lambda=(\ell-1)(\ell+2)/2$, and $f=1-2M/r$. The source term $\mathcal{S}$ encapsulates information about possible matter exciting the fluctuations $\Phi$. For other massless fields, a similar wave equation describes all radiative degrees of freedom, with an effective potential $V_{\rm RW}=f\left[\frac{\ell(\ell+1)}{r^2}-\frac{2(1-s^2)M}{r^3}\right]$, with $s=0,1$ for scalar or electromagnetic fields, respectively. We will use scalar fields as toy models in some of our discussions.
	
	In the absence of a source, the late time decay of generic massless fluctuations at fixed spatial position and asymptotically late times, corresponding to solutions of the linear wave equation, is governed by Price's law~\cite{Price:1971fb,Leaver:1986gd,Gundlach:1993tp,Hintz:2020roc},
	\be
	\Phi \sim t^{-p}\,.
	\ee
	This behavior is dictated by a branch cut at zero frequency~\cite{Leaver:1986gd,Ching:1994bd,Ching:1995tj}. For generic initial data ($\Phi(t=0)\neq0, \partial_t\Phi (t=0) \neq0 $) of compact support then $p=2\ell+3$. If initial data is of compact support and initially static ($\partial_t\Phi (t=0) =0 $) then $p=2\ell+4$~\cite{Karkowski:2003et,Price:2004mm}.
	For static initial data but non-compact support $p=2\ell+2$.
	As a special example, a late-time decay $t^{-3}$, was proven to describe the asymptotics of a self-gravitating scalar field, with compact-support initial conditions~\cite{Dafermos:2003yw}.
	
	In the presence of forcing terms, like matter moving in the BH exterior, the source $\mathcal{S}$ in \eqref{eq:massless} is nonzero. Its effect on possible late-time behavior has hardly been explored. However, when the full set of Einstein equations is expanded to second order around a fixed BH background, then second order metric quantities obey a similar equation, sourced by first order quantities. There are arguments suggesting that these lead to nontrivial power-law decay at late times~\cite{Okuzumi:2008ej}. Thus, here we focus on precisely the class of equations \eqref{eq:massless}, in BH spacetimes. 
	
	\section{Analytical predictions\label{app:analytical}}
	Here, we provide details on the derivation of the IDTs~(6) and SoDTs~(8)--(10), building on Refs.~\cite{Barack:1998bw,Karkowski:2003et,Price:2004mm}. 
	\subsection{IDTs: $(\varepsilon,\alpha)=(1,0)$}
	Equation~(1) at this order becomes
	\begin{equation}
		-4\partial_{uv}^2\Phi^{(1,0)}=V_0\Phi^{(1,0)}+\delta V \Phi^{(0,0)}.
	\end{equation}
	To set up the initial value problem, we specify initial conditions. We specifically consider initial data,
	\begin{equation}
		\begin{cases}
			&\Phi^{(0,0)}(u=u_I)=0\\
			&\Phi^{(0,0)}(v=0)=\Upsilon(u),
		\end{cases}\label{eq:initialdata}
	\end{equation}
	where $\Upsilon(u)$ is a function outside $x=x_0$, which has support only in narrow ranges spanned from $u=u_I$. The solution~$\Phi^{(1,0)}$ at ${\cal I}^+$ is then formally written as
	\beq
	\Phi_{{\cal I}^+}^{(1,0)}\left(u\right)&=&-\frac{1}{4}\int_{u_I}^{u}\mathrm{d}u'\int_{0}^\infty \mathrm{d}v'G^\infty\left(u;u',v'\right)\nonumber\\
	&\times&\delta V\left(u',v'\right)\Phi^{(0,0)}\left(u',v'\right), 
	\eeq
	where $\Phi_{\cal I^+} =\Phi(u, v\to\infty)$ is the value of the field at~$\cal I^+$; $G^\infty(u;u',v')$ is the (retarded) Green's function at ${\cal I}^+$ for the operator~$-4\partial_{uv}^2-V_0$. 
	
	Now, we separate the domain of the integration as
	\begin{equation}
		\begin{aligned}
			\label{eq:IntegralDecomposition}
			\int_{u_I}^{u}\mathrm{d}u'\int_{0}^\infty \mathrm{d}v'&=\int_{u_I}^{u}\mathrm{d}u'\int_{u+2x_0}^\infty \mathrm{d}v'\\
			&+\int_{u_I}^{v'-2x_0}\mathrm{d}u'\int_{0}^{u+2x_0} \mathrm{d}v'\\
			&+\int_{v'-2x_0}^u\mathrm{d}u'\int_{0}^{u+2x_0}\mathrm{d}v'.
		\end{aligned}
	\end{equation}
	As shown in~\cite{Barack:1998bw}, the contribution on the first line on the right-hand side is significant for the late-time behavior at~${\cal I}^+$; other contributions lead to an exponential decaying in $u$ or a subleading power-law decay. The domain of the main contribution for the late-time behavior mainly includes an asymptotically flat region.
	
	We henceforth focus on the main contribution for the late-time behaviors at ${\cal I}^+$,
	\beq
	\Phi_{{\cal I}^+}^{(1,0)}\left(u\right)&=&-\frac{1}{4}\int_{u_I}^{u}\mathrm{d}u'\int_{u+2x_0}^{\infty} \mathrm{d}v'G\left(u;u',v'\right)\nonumber\\
	&\times&\delta V\left(u',v'\right)\Phi^{(0,0)}\left(u',v'\right), \label{Eq:Phi10}
	\eeq
	where the Green's function at large distances~$(x\gg x_0)$ is given by~\cite{Barack:1998bw}\footnote{To obtain it from Eq.~(28) in \cite{Barack:1998bw}, we used
		\begin{equation}
			\begin{aligned}
				\frac{\partial^\ell}{\partial u^\ell}\left[\left(v'-u\right)^\ell \left(u-u'\right)^\ell\right]=&\ell!  \frac{\Gamma\left(2\ell+1\right)\Gamma\left(-2\ell\right)}{\Gamma\left(\ell+1\right)^2}\left(v'-u\right)^\ell\\
				&~\!_2{\bf F}_1\left(-\ell,-\ell;-2\ell;\frac{u'-v'}{u-v'}\right)\,,
			\end{aligned}
		\end{equation}
		where $~\!_2{\bf F}_1(a,b;c;z)$ is a regularized Gaussian hypergeometric function. Moreover, we made use of the definition of the hypergeometric series.
	}
	\begin{equation}
		\begin{aligned}
			\label{eq:Greenfunction}  
			G\left(u;u',v'\right)=&\sum_{j=0}^\ell \alpha_j \frac{(v'-u)^{\ell-j}}{(v'-u')^{\ell-j}}\,,
		\end{aligned}
	\end{equation}
	where we have introduced 
	\begin{equation}
		\label{eq;alphaj}
		\alpha_j\equiv \frac{\left(-1\right)^j}{j!}\frac{\Gamma\left(2\ell-j+1\right)}{\Gamma\left(\ell-j+1\right)^2}\,,
	\end{equation}
	where $\Gamma(z)$ is the Gamma function.
	We also have~\cite{Barack:1998bw}
	\begin{equation}
		\Phi^{(0,0)}\left(u,v\right)=\sum_{n=0}^\ell \frac{\left(2\ell-n\right)!}{n!\left(\ell-n\right)!} \frac{g^{(n)}\left(u\right)}{\left(v-u\right)^{\ell-n}}\,,
	\end{equation}
	where $g(u)$ is associated with the initial data~\eqref{eq:initialdata} by
	\begin{equation}
		\label{eq:functiong}
		g(u)=\frac{1}{(\ell-1)!}\int_{u_I}^u \mathrm{d}u'\left(\frac{u}{u'}\right)^{\ell+1}\left(u-u'\right)^{\ell-1}\Upsilon(u')\,.
	\end{equation}
	Here, the superscript~$(n)$ denotes a $n$-th order derivative.
	Equation~\eqref{Eq:Phi10} is rewritten as
	\begin{equation}
		\begin{aligned}
			\Phi_{{\cal I}^+}^{(1,0)}\left(u\right)&=-\frac{1}{4}\sum_{n=0}^\ell \frac{\left(2\ell-n\right)!}{n!\left(\ell-n\right)!}\int_{u_I}^u\mathrm{d}u'\\
			&\times \int_{u+2x_0}^\infty \mathrm{d}v'\frac{G(u;u',v')\delta V }{\left(v'-u'\right)^{\ell-n}}g^{(n)}\left(u'\right).
		\end{aligned}
	\end{equation}
	Integrating by part $n$ times over $u'$, we obtain
	\begin{equation}
		\begin{aligned}
			\label{eq:generalPhi10}
			\Phi_{{\cal I}^+}^{(1,0)}\left(u\right)&=-\frac{1}{4}\sum_{n=0}^\ell \frac{\left(2\ell-n\right)!}{n!\left(\ell-n\right)!}\int_{u_I}^u\mathrm{d}u'g\left(u'\right)
			\\
			&\times\int_{u+2x_0}^\infty \mathrm{d}v'\frac{\partial^n}{\partial u'^{n}}\left[\frac{G(u;u',v')\delta V }{\left(v'-u'\right)^{\ell-n}}\right].
		\end{aligned}
	\end{equation}
	Here, we have used the fact that $g(u')$, which is given by Eq.~\eqref{eq:functiong}, decays sufficiently fast at late times for compactly supported initial data~\eqref{eq:initialdata}.
	
	We have the relation
	\begin{equation}
		\sum_{n=0}^\ell \frac{\partial^n}{\partial u'^{n}}\left[\frac{G(u;u',v')\delta V }{\left(v'-u'\right)^{\ell-n}}\right]=\sum_{n=0}^\ell \sum_{j=0}^\ell {\cal P}_{j,n}^\ell\frac{\left(v'-u'\right)^{j-\rho-2\ell}}{\left(v'-u\right)^{j-\ell}}\,,
	\end{equation}
	where ${\cal P}_{j,n}^\ell$ is a constant labelled by $j$, $n$, $\ell$. Then, Eq.~\eqref{eq:generalPhi10} is rewritten as
	\begin{equation}
		\begin{aligned}
			\Phi_{{\cal I}^+}^{(1,0)}\left(u\right)&=-\frac{1}{4}\sum_{n=0}^\ell \sum_{j=0}^\ell \frac{\left(2\ell-n\right)!}{n!\left(\ell-n\right)!} {\cal P}_{j,n}^\ell\\
			&\times\int_{u_I}^u\mathrm{d}u'g\left(u'\right)\int_{u+2x_0}^\infty \mathrm{d}v'\frac{\left(v'-u'\right)^{j-\rho-2\ell}}{\left(v'-u\right)^{j-\ell}}\,.
		\end{aligned}
	\end{equation}
	Integrating it over $v'$, we obtain
	\begin{equation}
		\begin{aligned}
			\Phi_{{\cal I}^+}^{(1,0)}\left(u\right)&=-\frac{1}{4}\sum_{n=0}^\ell \sum_{j=0}^\ell\sum_{m=0}^{\ell-j} {\cal Q}_{j,n,m}^\ell u^{1-2\ell-\rho+j+m}\\
			&\times\int_{u_I}^u\mathrm{d}u'g\left(u'\right)\left(1+\frac{2x_0}{u}-\frac{u'}{u}\right)^{1-2\ell-\rho+j+m},
			\label{eq:Phiscrip10full}
		\end{aligned}
	\end{equation}
	where ${\cal Q}_{j,n,m}^\ell$ is a constant labelled by $j$, $n$, $\ell$, and $m$. Since $g(u)$ is compactly supported around $u=u_I$ and decays sufficiently fast at large~$u$, the leading contribution from the integration over $u'$ is constant at leading order in $1/u$. Therefore, the leading contribution in the sum in Eq.~\eqref{eq:Phiscrip10full} at large~$u$ comes from $m=\ell-j$ and takes the form,
	\begin{equation}
		\Phi_{\cal I^+}^{(1,0)}(u) \sim u^{1-\rho-\ell},~~
		\Phi^{(1,0)} \sim t^{-\rho-2\ell},~~(t\gg x)\,.\label{eq:IDTsinApp}
	\end{equation}
	Note that the other terms in the sum in Eq.~\eqref{eq:Phiscrip10full} are subleading to the contribution at $m=\ell-j$.
	Equation~\eqref{eq:IDTsinApp} corresponds to Eq.~(6). The case of $\rho=3$ precisely recovers Price's law~\cite{Price:1971fb}.
	Indeed, this corresponds to the first curvature correction that appears in the potential, $\delta V \sim r^{-3}$, leading to the slowest decaying tail.
	
	To summarize, the late-time behavior of $\Phi^{(1,0)}$ is described by the power-law tail~\eqref{eq:IDTsinApp}. The tails are generated in an asymptotically flat region and come from back-scattering of waves due to the weak curvature. The power is completely determined by the spacetime structure of backgrounds in an asymptotically flat region and the angular number of the waves.
	
	\subsection{SoDTs: $(\varepsilon,\alpha)=(0,1)$}
	Equation~(1) at this order reduces to 
	\begin{equation}
		\label{eq:SoDTeq}
		-4\partial_{uv}^2\Phi^{(0,1)}=V_0\Phi^{(0,1)}+{\cal S}.
	\end{equation}
	The solution at large distances~$(x\gg x_0)$ can be formally written as 
	\begin{equation}
		\begin{aligned}
			\Phi^{(0,1)}\left(u,v\right)&=-\frac{1}{4}\int_{-\infty}^{u}\mathrm{d}u'\\
			&\times \int_{-\infty}^v \mathrm{d}v'G\left(u,v;u',v'\right){\cal S}\left(u',v'\right)\,, 
		\end{aligned}
	\end{equation}
	where $G(u,v;u',v')$ is the (retarded) Green's function at $x\gg x_0$ for the operator~$-4\partial_{uv}^2-V_0$.  Since we are interested in the late-time behaviors measured at large distances, we focus on the domain that includes $x\ge x_0$ and very late times~$t\gg x$. Therefore, we here consider
	\begin{equation}
		\begin{aligned}
			\label{eq:phi01}
			\Phi^{(0,1)}\left(u,v\right)&=-\frac{1}{4}\int_{-\infty}^{u}\mathrm{d}u'\\
			&\times \int_{u+2x_0}^v \mathrm{d}v'G\left(u;u',v'\right){\cal S}\left(u',v'\right)\,, 
		\end{aligned}
	\end{equation}
	where $G\left(u;u',v'\right)$ is given by Eq.~\eqref{eq:Greenfunction}.
	In the following, we study two types of sources~${\cal S}$. One is a localized source, modeling a pointlike particle moving around a Schwarzschild BH. Another is an extended source, which mimics first-order sources in the context of second-order perturbation theory. 
	
	\subsubsection{Localized sources: timelike case~$(|U_s|<1)$}
	Consider the source,
	\begin{equation}
		\mathcal{S} = \frac{\delta(x-x_s-U_s t)}{x^\beta},~~\beta\ge 0\,,\label{localisedsources2}
	\end{equation}
	where $U_s$ is the velocity of the source. We assume $|U_s|<1$. 
	
	Consider first the late-time behavior at ${\cal I}^+$ when $U_s<0$. The source-driven contribution~\eqref{eq:phi01} is
	\begin{equation} \begin{aligned}
			\Phi_{{\cal I^+},-1<U_s<0}^{(0,1)}
			&=-\sum_{j=0}^\ell \alpha_j\frac{\left(1-U_s\right)^{\beta-1}\left(1+U_s\right)^{\ell-j}}{2^{1+\ell-j}U_s^{\beta+\ell-j}}I_{u',|U_s|\neq 1}^j,\\ 
			I_{u',|U_s|\neq 1}^j&\equiv \int_{-\infty}^{u}\mathrm{d}u'\left(u'-\frac{1-U_s}{1+U_s}u+\frac{2x_s}{1+U_s}\right)^{\ell-j}\\
			&\times\left(u'+\frac{x_s}{U_s }\right)^{-\beta-\ell+j}\\
			&\times H\left(\frac{1+U_s}{1-U_s}u'-u-2x_0+\frac{2x_s}{1-U_s}\right)\,.
			\label{eq:Iuvsneq1}
		\end{aligned}
	\end{equation}
	Here, $H$ is the Heaviside step function. Note that $\alpha_j$ is given by Eq.~\eqref{eq;alphaj}. The Heaviside step function in $I_{u',|U_s|\neq 1}^j$ in Eq.~\eqref{eq:Iuvsneq1} sets the lower limit of integration,
	\begin{equation}
		\label{eq:lowerup}
		u_{\rm lower}'\equiv \frac{1-U_s}{1+U_s}\left(u+2x_0-\frac{2x_s}{1-U_s}\right)\,.
	\end{equation}
	The causality condition for the retarded Green's function requires $u>u_{\rm lower}'$, or equivalently to
	\begin{equation}\label{eq:no-inwards-tails}
		U_s>-\frac{x_s-x_0}{u+x_0}\,,
	\end{equation}
	which reduces to $U_s>0$ at very late times~$t\gg x$; otherwise, $I_{u',|U_s|\neq 1}^j=0$. Therefore, \textit{inward-directed} localized sources have no contribution into the late-time behavior observed at ${\cal I}^+$. 
	
	\begin{figure}[h!]
		\includegraphics[width=0.7\linewidth]{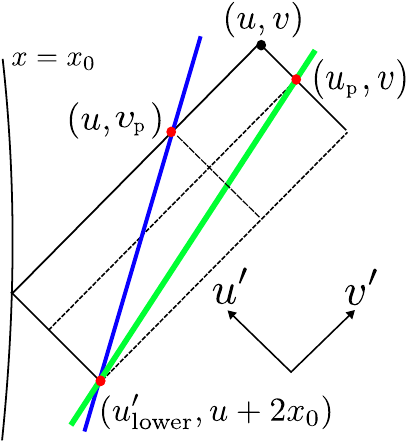}    
		\caption{Green (blue) solid lines denote trajectories of the outwards-directed sources that have (have not) already passed through the observer's location, respectively. The two sources have different velocities in this example. Expressions for $u_{\rm lower}'$, $v_{\rm p}$, and $u_{\rm p}$ are given by Eqs.~\eqref{eq:lowerup},~\eqref{eq:vp}, and~\eqref{eq:upperup}, respectively.} 
		\label{fig:spacetimediagram}
	\end{figure}
	Next, we discuss the late-time behavior in the case of $U_s>0$. In this case, the outwards-directed source passes through the observer's location at
	\begin{equation}
		\begin{aligned}
			\label{eq:tp}
			t=& t_{\rm p}\,, \quad t_{\rm p}\equiv\frac{x-x_s}{U_s}\,,
		\end{aligned}
	\end{equation}
	in $(t,x)$ coordinates, which is equivalent to
	\begin{equation}
		\begin{aligned}
			\label{eq:vp}
			v=&v_{\rm p}\,, \quad v_{\rm p}\equiv \frac{1+U_s}{1-U_s}u+\frac{2x_s}{1-U_s}\,,
		\end{aligned}
	\end{equation}
	or
	\begin{equation}
		\begin{aligned}
			\label{eq:upperup}
			u=u_{\rm p}\,, \quad u_{\rm p}\equiv \frac{1-U_s}{1+U_s}v-\frac{2x_s}{1-U_s}\,.
		\end{aligned}
	\end{equation}
	The trajectories of sources that have already passed or not through the observer can be understood in Fig.~\ref{fig:spacetimediagram}. 
	
	The source-driven contribution~\eqref{eq:phi01} {\it before} passing~($v>v_{\rm p}$ or $u<u_{\rm p}$) becomes
	\begin{equation} \begin{aligned}
			\Phi_{\mathrm{b},0<U_s<1}^{(0,1)}
			=&-\sum_{j=0}^\ell \alpha_j\frac{\left(1-U_s\right)^{\beta-1}\left(1+U_s\right)^{\ell-j}}{2^{1+\ell-j}U_s^{\beta+\ell-j}}I_{u',|U_s|\neq 1}^j.
		\end{aligned}
	\end{equation}
	Here, the subscript~``$\mathrm{b}$'' denotes {\it before} passing. Note that $I_{u',|U_s|\neq 1}^j$ is given by Eq.~\eqref{eq:Iuvsneq1}. Since the inequality~\eqref{eq:no-inwards-tails} is satisfied for the outwards-directed sources, $I_{u',|U_s|\neq 1}^j$ has non-trivial contributions into the late-time behavior. Integrating it over $u'$, we obtain,
	\beq
	\begin{aligned}
		&\Phi_{\mathrm{b},0<U_s<1}^{(0,1)}=\sum_{j=0}^\ell c_{\mathrm{b},1}^j \left(u+\frac{x_s}{U_s}\right)^{1-\beta}\mkern-12mu _2F_1\left({\frak a},{\frak b};{\frak c};\frac{2U_s}{U_s-1}\right)\\
		&\!+c_{\mathrm{b},2}^j \left(u+\frac{x_s}{U_s}\right)^{-\beta-\ell+j}\mkern-12mu _2F_1\left({\frak a},{\frak b};{\frak c};-\frac{2U_s x_0}{u U_s+x_s}\right)\,,\label{eq:tailsbeforejump}
	\end{aligned}
	\eeq
	where $c_{\mathrm{b},1}^j$ and $c_{\mathrm{b},2}^j$ are non-vanishing constants; ${}_2F_1({\frak a},{\frak{b};{\frak c};z})$ is a Gaussian hypergeometric function. We defined
	\begin{equation}
		\begin{aligned}
			{\frak a}=&1+\ell-j,~~{\frak b}=&\beta+\ell-j,~~{\frak c}=&2+\ell-j\,.
		\end{aligned}
	\end{equation}
	
	The source-driven contribution {\it after} passage~($v<v_{\rm p}$ or $u>u_{\rm p}$) takes the form,
	\begin{equation} \begin{aligned}
			\Phi_{\mathrm{a},0<U_s<1}^{(0,1)}
			&=-\sum_{j=0}^\ell \alpha_j\frac{\left(1-U_s\right)^{\beta-1}\left(1+U_s\right)^{\ell-j}}{2^{1+\ell-j}U_s^{\beta+\ell-j}}I_{u',\mathrm{a}}^j,\\
			I_{u',\mathrm{a}}^j&\equiv \int_{-\infty}^{u}\mathrm{d}u'\left(u'-\frac{1-U_s}{1+U_s}u+\frac{2x_s}{1+U_s}\right)^{\ell-j}\\
			&\times\left(u'+\frac{x_s}{U_s }\right)^{-\beta-\ell+j}\\
			&\times H\left(\frac{1+U_s}{1-U_s}u'-u-2x_0+\frac{2x_s}{1-U_s}\right)\\
			&\times H\left(-\frac{1+U_s}{1-U_s}u'+v-\frac{2x_s}{1-U_s}\right).
		\end{aligned}
	\end{equation}
	Here, the subscript~``$\mathrm{a}$'' denotes {\it after} passing. The first step function on the right-hand side in $I_{u',a}^j$ sets the lower limit on the integration, $u_{\rm lower}'$ in Eq.~\eqref{eq:lowerup}; the second step function sets the upper limit, $u_{\rm p}$ in Eq.~\eqref{eq:upperup}. Integrating it over $u'$, we obtain
	\beq
	\begin{aligned}
		&\Phi_{\mathrm{a},0<U_s<1}^{(0,1)}=\sum_{j=0}^\ell c_{\mathrm{a},1}^j \left(u+\frac{x_s}{U_s}\right)^{-\beta-\ell+j}\left(u-v\right)^{1+\ell-j}\\
		&\times~_2F_1\left({\frak a},{\frak b};{\frak c};\frac{\left(u-v\right)U_s}{u U_s+x_s}\right)\\
		&+c_{\mathrm{a},2}^j \left(u+\frac{x_s}{U_s}\right)^{-\beta-\ell+j}\!_2F_1\left({\frak a},{\frak b};{\frak c};-\frac{2U_s x_0}{u U_s+x_s}\right),\label{eq:tailsafterjump}
	\end{aligned}
	\eeq
	where $c_{\mathrm{a},1}^j$ and $c_{\mathrm{a},2}^j$ are non-vanishing constants. 
	Noting that there is an upper bound on $v$~(currently $v<v_{\rm p}$), we obtain the leading asymptotic behavior at $v\sim v_{\rm p}$ with large values,
	\begin{equation} 
		\begin{aligned}
			\Phi_{\mathrm{a},0<U_s<1}^{(0,1)}&\sim u^{1-\beta}\sum_{j=0}^\ell c_1^j~_2F_1\left({\frak a},{\frak b};{\frak c};\frac{2U_s}{U_s-1}\right)\\
			&+c_2^ju^{-1-\ell+j}\,.
		\end{aligned}
	\end{equation}
	Equations~\eqref{eq:tailsbeforejump} and~\eqref{eq:tailsafterjump} agree each other at $v=v_{\rm p}$ and $u=u_{\rm p}$. Comparing them, one can see that the behavior of $\Phi^{(0,1)}$ changes before and after the source passes through the observer's location. This suggests that one can observe a ``jump'' of the tail-exponent before and after passing. This is indeed confirmed in the numerics~(see the jump around $t/M=360$ in the bottom figure in Fig.~1.
	
	To summarize, outward-directed sources with $0<U_s< 1$ cause SoDTs. After passing, we observe at late times~(see also footnote~2):
	\begin{equation}
		\Phi_{{\cal I^+},0<U_s< 1}^{(0,1)}\sim u^{1-\beta},~~
		\Phi_{0<U_s<1}^{(0,1)}\sim t^{-(\beta+\ell)},~~(t\gg x)\,.\label{eq:LSoDTsinApp}
	\end{equation}
	These correspond to Eq.~(8).

	The origin of the tail phenomena above is fundamentally different from tails generated by back-scattering due to the weak curvature in an asymptotically flat region. 
	In fact, at this order in the perturbative expansion~(5), we are considering the wave propagation problem in flat space, and still obtaining power-law tails. 
	The SoDTs~\eqref{eq:LSoDTsinApp} are generated in flat space due to the presence of outward-directed sources. The exponent is determined by $\beta$ and an angular number of waves.
	
	\subsubsection{Localized sources: null case~($|U_s|=1$)}
	Consider now the source \eqref{localisedsources2} with $U_s=\pm1$, and focus on the late-time behavior of $\Phi^{(0,1)}$ at ${\cal I}^+$. In the following, although we implicitly assume $\beta\ge2$, the conclusion includes the case of $\beta=0,1$ as well. The source-driven contribution~\eqref{eq:phi01} is then explicitly written as
	\begin{equation} 
		\begin{aligned}
			\Phi_{{\cal I^+},U_s=-1}^{(0,1)}&=-\sum_{j=0}^\ell2^{-2+\beta}\alpha_j H\left(-u-2x_0+x_s\right)\\
			&\times \int_{-\infty}^u \mathrm{d}u'\frac{\left(-u+x_s\right)^{\ell-j}}{\left(-u'+x_s\right)^{\ell-j+\beta}}\,,
		\end{aligned}
	\end{equation}
	for $U_s=-1$, and
	\begin{equation} \begin{aligned}
			\Phi_{{\cal I^+},U_s=1}^{(0,1)} 
			=&-\sum_{j=0}^\ell\sum_{k=0}^{\ell-j} 2^{-2+\beta}\left(2x_0\right)^{\ell-k-j}I_{u',U_s=1}^{j,k}\,,
		\end{aligned}
	\end{equation}
	for $U_s=1$. Here, we have introduced
	\beq
	\begin{aligned}
		I_{u',U_s=1}^{j,k}&\equiv c^{j,k}\int_{-\infty}^{u}\mathrm{d}u' \left(u-u'+2x_0\right)^{1-\beta-\ell+j+k}\\
		&\times \delta\left(u'+x_s\right)\,, \quad \text{with}\\
		c^{j,k}&\equiv \alpha_j \frac{\Gamma\left(\ell-j+1\right)\Gamma\left(\ell-j+\beta-k-1\right)}{\Gamma\left(\ell-j+\beta\right)\Gamma\left(\ell-j-k+1\right)}\,,\label{eq:Iuvs1} 
	\end{aligned}
	\eeq
	and $\alpha_j$ is given by Eq.~\eqref{eq;alphaj}. Note that $\sum_{j=0}^\ell c^{j,\ell-j}=0$ for $2\le \beta\le \ell+1$ for $\ell\ge 1$. Thus, the contribution $\Phi_{{\cal I^+},U_s=-1}^{(0,1)}$ contains a~$H(-u-2x_0+x_s)$ term: inward-directed sources with $U_s=-1$ have no contribution to SoDTs in $u>x_s-2x_0$.
	The integration~$I_{u',U_s=1}^{j,k}$ in Eq.~\eqref{eq:Iuvs1} then yields
	\begin{equation}
		\begin{aligned}
			I_{u',U_s=1}^{j,k}=c^{j,k} u^{1-\beta-\ell+j+k}\left(1+\frac{2x_0}{u}+\frac{x_s}{u}\right)^{1-\beta-\ell+j+k}\,.
		\end{aligned}
	\end{equation}
	When performing the summation of $I_{u',U_s=1}^{j,k}$ over $j$ and $k$, one can use again that $\sum_{j=0}^\ell c^{j,\ell-j}=0$ when $2\le \beta\le \ell+1$ for $\ell\ge 1$. Therefore, the leading term in the sum at large~$u$ comes from~i) $k=\ell-j$ when $\beta$ is generic for $\ell=0$ and when $\beta=0,1$ or $\beta\ge \ell+2$ for $\ell \ge 1$;~ii) $k=\ell-j-1$ when $2\le \beta \le \ell+1$ for $\ell\ge1$. 
	
	We thus show that in the presence of {\it outward-directed sources with $U_s=1$}, the source-driven contribution~$\Phi^{(1,0)}$ has power-law tails,
	\begin{equation}
		\label{eq:pforLocalisedNull}
		\Phi_{{\cal I^+},U_s=1}^{(0,1)} \sim u^{1-\beta},~~\Phi_{U_s=1}^{(0,1)}\sim t^{-\beta-\ell},~~(t\gg x)\,,
	\end{equation}
	for any $\beta$ when $\ell=0$, when $\beta=0,1$ or when $\beta \ge \ell+2$ for $\ell\ge1$. In addition, we find
	\begin{equation}
		\label{eq:pforLocalisedNull2}
		\Phi_{{\cal I^+},U_s=1}^{(0,1)} \sim u^{-\beta},~~\Phi_{U_s=1}^{(0,1)}\sim t^{-\beta-\ell-1},~~(t\gg x)\,,
	\end{equation}
	when $2\le \beta \le \ell+1$ for $\ell\ge 1$. These correspond to Eqs.~(9) and~(10). The exponent is thus completely determined by $\beta$ and the angular number of the waves. 
	
	\subsubsection{Extended sources}
	Within second-order perturbation theory, we consider source terms describing contributions from first-order perturbations, which we take to have the form,
	\begin{equation}
		\label{eq:extendedsourse}
		{\cal S}=\frac{\Psi_2}{x^\beta} H\left(u-u_s\right),~~\beta\ge 2\,,
	\end{equation}
	where
	\begin{equation}
		\Psi_2\equiv \left(C_{q1} e^{-i \omega_1 u}+C_{t1}u^{-p_1}\right)\left(C_{q2} e^{-i \omega_2 u}+C_{t2}u^{-p_2}\right)\,.\nonumber
	\end{equation}
	We introduced  constants~$C_{q1},C_{q2},C_{t1},C_{t2}$, complex frequencies~$\omega_1,\omega_2$ whose imaginary parts are negative, and integers~$p_1, p_2$ such that $p_1,p_2\ge 2$. To analyze the source-driven term~\eqref{eq:phi01} with source~\eqref{eq:extendedsourse}, consider
	\begin{equation}
		{\cal S}=x^{-\beta}\left(C_qe^{-i\omega u}+C_t u^{-p}+C_{qt}e^{-i\tilde{\omega}u} u^{-\tilde{p}}\right)H\left(u-u_s\right),\nonumber
	\end{equation}
	where $C_q,C_t,C_{qt}$ are constants, $\omega=\omega_1+\omega_2$, $p=p_1+p_2$; $(\tilde{\omega},\tilde{p})$ corresponds to $(\omega_1,p_2)$ or $(\omega_2,p_1)$. The source-driven term~\eqref{eq:phi01} in this case becomes
	\beq
	\begin{aligned}
		&\Phi_{{\cal I}^+}^{(0,1)}\left(u\right)=2^{\beta-2}\sum_{m=0}^\ell\alpha_m \int_{u_s}^{u}\mathrm{d}u'\int_{u+2x_0}^\infty \mathrm{d}v'\\ &\frac{\left(v'-u\right)^{\ell-m}}{\left( v'-u'\right)^{\ell+\beta-m}} \left(-C_qe^{-i\omega u'}-C_t u'^{-p}-C_{qt}e^{-i\tilde{\omega}u'} u'^{-\tilde{p}}\right).
	\end{aligned}
	\eeq
	Integrating it over $v'$, we obtain
	\beq
	\mkern-32mu\Phi_{{\cal I}^+}^{(0,1)}\left(u\right)=-\sum_{m=0}^\ell\sum_{j=0}^{\ell-m} 2^{\beta+\ell-m-j-2}x_0^{\ell-m-j}I_{u'}^{m,j}\,,
	\eeq
	where we define
	\beq
	\begin{aligned}
		I_{u'}^{m,j}&\equiv c^{m,j} \int_{u_s}^{u}\mathrm{d}u'\Bigl(C_qe^{-i\omega u'}+C_t u'^{-p}+C_{qt}e^{-i\tilde{\omega}u'} u'^{-\tilde{p}}\Bigr)\\\nonumber
		&\times u^{1-\beta+j+m-\ell}\left(1+\frac{2x_0}{u}-\frac{u'}{u}\right)^{1-\beta+j+m-\ell}\,,
	\end{aligned}
	\eeq
	where $c^{m,j}$ is given in Eq.~\eqref{eq:Iuvs1}. As already noted below Eq.~\eqref{eq:Iuvs1}, $\sum_{m=0}^\ell c^{m,\ell-m}=0$ for $2\le \beta\le \ell+1$ for $\ell\ge 1$.
	
	For negative ${\rm Im}~\omega$ and ${\rm Im}~\tilde{\omega}$, the integrand in $I_{u'}^{m,j}$ has a peak around $u'=0$ and exhibits a damped oscillation as $u'$ is increased. This means that the contribution into the integral over $u'$ mainly comes from the domain~$u'\ll u$. Therefore, $I_{u'}^{m,j}$ at late times is well approximated by
	\beq
	\begin{aligned}
		&I_{u'}^{m,j}\simeq c^{m,j}  u^{1-\beta+j+m-\ell}\int_{u_s}^{u}\mathrm{d}u'\\
		&\times\left(C_qe^{-i\omega u'}+C_t u'^{-p}+C_{qt}e^{-i\tilde{\omega}u'} u'^{-\tilde{p}}\right)\,.
	\end{aligned}
	\eeq
	Performing the integration, we obtain
	\beq
	\begin{aligned}
		I_{u'}^{m,j}&\propto c^{m,j} u^{1-\beta+j+m-\ell}\Bigl{[}C_q\frac{i}{\omega}e^{-i u_s\omega}\\&+C_t\frac{u_s^{1-p}}{1-p}
		-C_{qt}u_s^{1-\tilde{p}}E_{\tilde{p}}\left(iu_s \tilde{\omega}\right)\Bigr{]}\,,
	\end{aligned}
	\eeq
	at leading order in $u$. Here, $E_{\tilde{p}}(z)$ is an exponential integral. Note that these come from the lower limit of the integration over $u'$, meaning the leading contribution comes from the intersection between the support of the Green's function and the source distribution. This is consistent with the observation in~\cite{Okuzumi:2008ej}.
	
	As before, $\sum_{m=0}^\ell c^{m,\ell-m}=0$ when $2\le \beta\le \ell+1$ for $\ell\ge 1$. Therefore, the leading term in the sum at large~$u$ comes from~i) $k=\ell-j$ when $\beta$ is generic for $\ell=0$, when $\beta=0,1$ or when $\beta\ge \ell+2$ for $\ell \ge 1$;~ii) $k=\ell-j-1$ when $2\le \beta \le \ell+1$ for $\ell\ge1$.
	
	Thus, we obtain the leading asymptotic behavior of $\Phi^{(0,1)}$ at late times:
	\begin{equation}
		\Phi_{{\cal I^+}}^{(0,1)} \sim u^{1-\beta},~~\Phi^{(0,1)}\sim t^{-\beta-\ell},~~(t\gg x)\,,
	\end{equation}
	for any $\beta$ when $\ell=0$, when $\beta = 0,1$ or when $\beta \ge \ell+2$ for $\ell\ge1$. Moreover, we find
	\begin{equation}
		\Phi_{{\cal I^+}}^{(0,1)} \sim u^{-\beta},~~\Phi^{(0,1)}\sim t^{-\beta-\ell-1},~~(t\gg x)\,,
	\end{equation}
	when $2\le \beta \le \ell+1$ for $\ell\ge 1$. This behavior is identical to that of localized outwards-directed sources case at the speed of light~(see Eqs.~\eqref{eq:pforLocalisedNull} and~\eqref{eq:pforLocalisedNull2}). The exponent is determined by the spatial distribution of the source and the angular properties of the wave. The SoDTs from extended sources can decay slower than IDTs~(6), suggesting that first-order perturbation theory fails to predict the late-time behaviors of waves at large distances. It is worth mentioning that the leading behavior arises from the lower limit of the integration over~$u'$, i.e., $u'=u_s$ hypersurface, suggesting that second-order tails are generated at the wavefront of first-order perturbations and propagate along the past light cone of distant observers. The first-order quasinormal modes can generate non-oscillating tails because they do not oscillate along $u'=u_s$ hypersurfaces.
	
	\section{Numerical procedure}
	We used different numerical routines to obtain the results in the main text, as a way to validate our results independently. Our first code is a time-domain solver with a two-step Lax-Wendroff integrator with second-order finite differences, which has been thoroughly described and validated in Refs.~\cite{Krivan_1997, Pazos_valos_2005, Zenginoglu:2011zz, Zenginoglu:2012us, Cardoso:2021vjq}. To achieve the necessary precision for the study of tails, we employ quadrupole floating point precision. Some of the results are extracted at null infinity, which is achieved by a hyperboloidal compactification of the spatial coordinate, with the outer boundary of the compactified domain corresponding to null infinity $\mathcal{I}^+$.
	
	Another code employs an 8th-order finite difference method for spatial discretization, the canonical Runge-Kutta 4th-order method for time integration and quadruple-precision floating-point. We use the outflow boundary condition as proposed by Ref.~\cite{Calabrese:2005fp}, placing it distant from the region where we extract the data to mitigate the impact of an imperfect boundary condition.
	
	We will often work with pointlike particles as our radiation source. They are represented by Dirac-delta distributions localized around the spatial location of the particle at a given instant, $x_{\rm p}(t)$. We approximate the Dirac-deltas with narrow Gaussian distributions  
	\beq
	\delta(x-x_{\rm p}(t)) =\frac {\exp\left[ -(x-x_{\rm p}(t))^2/(2\lambda_{\rm p}^2) \right]}{\sqrt{2\pi}\lambda_{\rm p}} \,,\label{delta_approximation}
	\eeq
	where $\lambda_{\rm p}$ is sufficiently small to ensure numerical convergence as $\lambda_{\rm p} \rightarrow 0$. We take $\lambda_{\rm p} \approx 4\mathrm{d}x$, where $\mathrm{d}x$ is the grid discretization step~\cite{Lopez-Aleman:2003sib}. Typically, we use $\mathrm{d}x \lesssim 0.05$.
	
	Throughout this work, we use three sets of initial data. (i) Static initial data (s), which corresponds to a Gaussian localized at some initial radius $x_I$
	\beq
	\mkern-22mu\phi_\text{s} (0, x) =\exp(- \frac{(x - x_I)^2}{2 \sigma^2})\,,\quad \partial_t \phi_\text{s} (0, x) = 0\,,\label{eq:ID_static}
	\eeq
	where $\sigma$ controls the width of the pulse. (ii) Non-static/generic (g) initial data of the form
	\beq
	\mkern-22mu\phi_\text{g}(0, x) =0 \,,\quad \partial_t \phi_\text{g}(0, x) = \exp(- \frac{(x- x_I)^2}{2 \sigma^2})\,.\label{eq:ID_generic} 
	\eeq
	(iii) Zero initial data for which $\phi(0, x) =\partial_t \phi(0, x) =0$. We have checked that all types of initial data listed above are in excellent agreement with the analytical predictions for tails of the homogeneous Zerilli equation. 
	
	The main quantity extracted from numerical simulations is the local decay rate $p=p(t)$, which we evaluate as
	\begin{equation}
		p= -t \partial_t \ln \abs{\phi}\,,\label{eq:p_def}
	\end{equation}
	such that $p$ is a constant for a field $\Phi\sim t^{-p}$.
	
	\section{Second order perturbations\label{app:secondorder}}
	%
	For the numerical simulation of the second order equation~(15), we use the hyperboloidal coordinates $\{\tau, \rho, \theta, \varphi\}$ with minimal gauge \cite{Ansorg:2016ztf, PanossoMacedo:2018hab, PanossoMacedo:2018gvw}, where the coordinate transformation from standard coordinates is given by 
	\begin{align}
		&\frac{t}{4M}=\tau-\frac{1}{2}\left(\ln \rho+\ln (1-\rho)-\frac{1}{\rho}\right)\,, \\
		&\frac{x}{4M}=\frac{1}{2}\left(\frac{1}{\rho}+\ln (1-\rho)-\ln \rho\right)\,.
	\end{align}
	We will use $p= -\tau \partial_\tau \ln \abs{\Psi}$ in this section.
	To allow a stable evolution with larger time-steps, we introduce a new momentum variable $\,^{\{n\}}\Pi = \partial_\tau \,^{\{n\}}\hspace{-.5mm}\Psi - (\rho - 1)\partial_\rho \,^{\{n\}}\hspace{-.5mm}\Psi$. We use the following types of initial data for first order polar perturbation with parameters $\rho_0 = 0.6$ and $\sigma^2 = 1 / 1000$:
	
	\noindent Approximately Ingoing:
	\beq
	\mkern-32mu&&\,^{\{1\}}\hspace{-.5mm}\Psi(\tau = 0, \rho) = \exp(-\frac{(\rho- \rho_0)^2}{2 \sigma^2}) \,, \\
	\mkern-32mu&&\,^{\{1\}}\hspace{-.5mm}\Pi(\tau = 0, \rho) = (\rho + 1)^{-1} \partial_\rho \,^{\{1\}}\Psi(\tau = 0, \rho)\,.\label{eq:id_2nd_order_ingoing}
	\eeq
	
	\noindent Approximately Outgoing:
	\beq
	&&\,^{\{1\}}\hspace{-.5mm}\Psi(\tau = 0, \rho) = \exp(-\frac{(\rho- \rho_0)^2}{2 \sigma^2}) \,, \\
	&&\,^{\{1\}}\hspace{-.5mm}\Pi(\tau = 0, \rho) = 0 \,.\label{eq:id_2nd_order_outgoing}
	\eeq
	For simplicity, we set the initial data of the second-order polar perturbation to zero, but we checked that its power law doesn't depend on the initial data.
	\be
	\,^{\{2\}}\hspace{-.5mm}\Psi(\tau = 0, \rho) = 0 \,, \qquad
	\,^{\{2\}}\hspace{-.5mm}\Pi(\tau = 0, \rho) = 0 \,.
	\ee
	To resolve the late-time tails, high-accuracy spatial discretization methods such as high-order finite differences or the pseudospectral method are required.
	Here, we use the Chebyshev pseudo-spectral methods for spatial discretization and the canonical Runge-Kutta 4th order method for time integration.
	High-precision floating-point numbers are also required to achieve the necessary precision for the study of late-time tails of second-order perturbation.
	For the final results in this section, we use $N=450$ for Chebyshev pseudo-spectral methods, and integrate the equation using a fixed step $\Delta \tau = 0.0005$, with the numerical precision being around 210 bits.
	\section{Tails of time derivatives}
	\begin{figure}[t!]
		\includegraphics[width=0.95\linewidth]{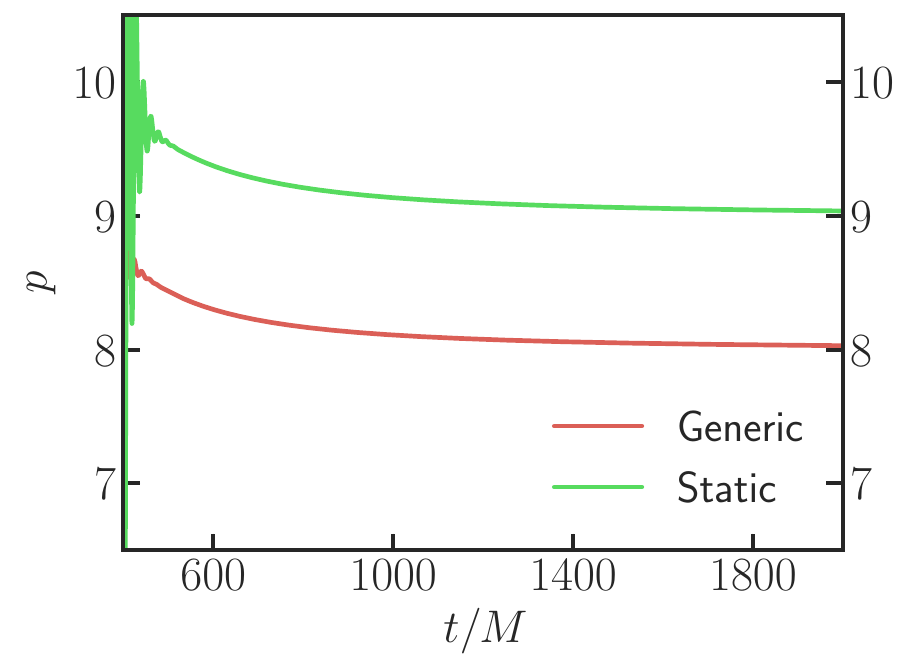}
		\caption{Power-law for the $\ell = 2$ solution of Eq.~\eqref{eqpsi2} with $\,^{\{2\}} \hspace{-.5mm}\dot{\mathcal{S}}_\Psi = 0$, using initial data from Eq.~\eqref{eq:X_initial_data}. Red line corresponds to using $\Psi_0$ as defined in Eq.~\eqref{eq:ID_generic}, green line to $\Psi_0$ as defined in Eq.~\eqref{eq:ID_static}. Data is extracted at $x = 100 M$.}
		\label{fig:second_order_pert_p_X}
	\end{figure}
	Suppose $\Psi$ is the Zerilli wavefunction, obeying Zerilli's equation~(15). Take a time derivative of that, to get
	\begin{align}
		-\partial_t^2 X+\partial_{x}^2 X-V_Z X&=\,^{\{2\}}\hspace{-.5mm}\dot{\mathcal{S}}_\Psi\,,\label{eqpsi2}
	\end{align}
	where we defined $X=\,^{\{2\}}
	\hspace{-.5mm}\dot{\Psi}$; dot is a time derivative. This can be done an arbitrary number of times, i.e., the $n-$th time derivative of the Zerilli wavefunction obeys the Zerilli equation. This looks like a contradiction: the $n$-th order derivative of the Zerilli wavefunction obeys the Zerilli equation, hence should have the same tails. However, the initial data is now
	\beq
	\,^{\{2\}}
	\hspace{-.5mm}\Psi(t=0)&=&\Psi_0(r)\,,\,^{\{2\}}
	\hspace{-.5mm}\dot{\Psi}(t=0)=\dot{\Psi}_0\,,\\
	X(t=0)&=&\dot{\Psi}_0\,,\dot{X}(t=0)=\ddot{\Psi}_0\,.
	\eeq
	Thus, there are constraints that need to be satisfied if $X$ is indeed the time derivative of the original master function. In particular, using the Zerilli equation we find
	\be
	X(t=0)=\dot{\Psi}_0\,,\qquad
	\dot{X}(t=0)=\partial^2_x\Psi_0-V_Z\Psi_0\,.\label{eq:X_initial_data}
	\ee
	Our numerical results confirm that initial conditions \eqref{eq:X_initial_data} are indeed ``special''. Based on Price's results one would expect naively that they would produce tails with power $p=2\ell +3$ for generic initial data and $2\ell+4$ for initially static data. We find, however---Fig.~\ref{fig:second_order_pert_p_X}---that $p=2\ell+4$ and $p=2\ell+5$ respectively. Thus, there {\it are} special classes of initial data that escape Price's description, as had to be the case for consistency.
	
	\bibliography{ref}

\begin{thebibliography}{68}
\bibitem [{com()}]{comment_exception}%
  \BibitemOpen
  \href@noop {} {}\bibinfo {note} {We find two curious exceptions to the
  empirical numerical results quoted. The first, for $U_s<1$, concerns $\ell =
  1,\,\beta = 3$ for which we find the power $p=5$, not $4$ as the empirical
  rule quoted suggests. The second concerns $U_s=1$ and refers to $\ell =
  0,\,\beta = 3$ for which our numerics give $p=4$, rather than the empirical
  rule~\eqref{eq:Static_v1}. These are the only known exceptions for $\ell \leq
  4$ and $\beta \leq 7$. Similarly, not all second order quantities seem to
  follow \eqref{eq:Static_v1} (in particular, the coupling of $\ell=m=2$ to
  $\ell=-m=2$).}\BibitemShut {Stop}%
\end{thebibliography}
\end{document}